\documentclass[11pt]{article}
\usepackage{amsfonts}
\usepackage{amscd}

\usepackage{bezier}

\newcommand{\sect}[1]{\setcounter{equation}{0}\section{#1}}
\newcommand{\subsect}[1]{\subsection{#1}}

\newcommand{\beq}{\begin{equation}}
\newcommand{\eeq}{\end{equation}}
\newcommand{\ben}{\begin{eqnarray}}
\newcommand{\een}{\end{eqnarray}}
\newcommand{\f}{\frac}

\newcommand{\ds}{\displaystyle}

\def\1{\'{\i}}                           
\def\R{{\mathbb R}} 
\def\>#1{{\bf #1}}                 
 
\def\d{{\rm d}}
\def\ie{{\rm i}}

\def\por{\!\cdot\!}

\begin{document}

\thispagestyle{empty}

\ 
\hfill\   hep-th/0305033

\bigskip 
\medskip

\begin{center}

{\Large{\bf{A new  `doubly special relativity' theory
from\\[4pt] a   quantum Weyl--Poincar\'e  algebra}}}

\end{center}

\bigskip 
\medskip

\begin{center}    Angel Ballesteros$^a$, N.~Rossano
Bruno$^{a,b}$ and Francisco~J.~Herranz$^a$
\end{center}

\begin{center} {\it { 
${}^a$Departamento de F\1sica, Universidad de Burgos, Pza.\
Misael Ba\~nuelos s.n., \\ 09001 Burgos, Spain }}\\ e-mail:
angelb@ubu.es, fjherranz@ubu.es
\end{center}

\begin{center} {\it { 
${}^b$Dipartimento di Fisica, Universit\`a di  Roma ROMA
TRE, and INFN Sez.\ Roma Tre, Via Vasca Navale 84, 00146
Roma, Italy}}\\ e-mail: rossano@fis.uniroma3.it
\end{center}

\bigskip 
\medskip

\begin{abstract}
\noindent A    `mass-like' quantum  Weyl--Poincar\'e 
algebra  is proposed to describe, after the identification
of the   deformation parameter with the Planck length, a
new relativistic theory  with two observer-independent
scales (or DSR theory). Deformed  momentum representation, 
finite boost transformations, range of rapidity, energy and
momentum, as well as position and velocity operators are
explicitly studied and compared with those of previous DSR
theories based on $\kappa$-Poincar\'e algebra. The main
novelties of the   DSR theory here presented are the new
features of momentum saturation and a new type of deformed
position operators.
\end{abstract}

\bigskip\medskip 
\bigskip\medskip

\newpage

%%%%%%%%%%%%%%%%%%%%%%%%%%%%%%%%%%%%%%%%%%%%%%%%
\sect{Introduction}

 In the last few years several approaches to the problem of
unification of general relativity and quantum mechanics
have led to arguments in favour of the modification of
Lorentz symmetry at the Planck scale \cite{Garay}. The
different approaches to quantum gravity such as loop
quantum gravity {\cite{Carlip:2001wq,Smolin:2002sz}}, or
string theory
\cite{Polchinski:1996na,Forste:2001ah} assigns to Planck
scale a fundamental role in the structure of spacetime  and
of momentum-space, or as linked to discrete spectra of
physical observables
\cite{Freidel:2002hx}.

The so called `doubly special relativity' (DSR) theories
 (see~\cite{Amelino-Camelia:2002vy} and references
therein)  have been proposed as possible tools to
investigate this ongoing quantum gravity debate in order to
reconcile Lorentz invariance with the new fundamental role
assigned to Planck scale.   Different DSR proposals
\cite{Amelino-Camelia:2000mn,MagueijoSmolin,
Amelino-Camelia:2002gv}
introduce   in addition to the usual observer-independent
velocity scale, an observer-independent length scale
(momentum scale), possibly related with the Planck scale
($\ds L_p\simeq 10^{-33}$ cm), which acquires the role of a
minimum length (maximum momentum)
\cite{MagueijoSmolin,Bruno:primo}, in a way in which Lorentz
invariance is preserved.  Also the observable implications
of DSR theories are being studied in connection with some
planned Lorentz-symmetry tests
\cite{grbgac,glast}, and in searching for a kinematical
solution for the puzzling observations of ultra-high-energy
cosmic rays
\cite{cosmicRAYdata,kifune,AmelinoPiranPRD}. 

Under investigation
\cite{Bruno:primo,Kowalski-Glikman:2002we,Lukierski:2002df}
is the role that quantum groups can play in these DSR
theories; for most of them,   the $\kappa$-Poincar\'e
algebra
 \cite{LukierskiRuegg1992,Majid:1994cy,lukbasis} seems to
play  a role which is analogous to that played by
Poincar\'e group in Einstein's special relativity. 
However, in pursuing this connection between relativistic
theories  with two observer-independent scales and quantum
groups, it is natural to consider that {\em any} quantum
Poincar\'e algebra should be taken as a first stage prior
to the search of  more general DSR theories which should be
endowed with either a quantum conformal symmetry or a 
$q$-AdS ($q$-dS) symmetry.  In this respect, several quantum
deformations for $so(p,q)$ algebras of non-standard or
twisted type (different from the Drinfeld--Jimbo
deformation) have been constructed during the last years.
Some of them have been obtained within a conformal
framework as
 quantum conformal algebras of the Minkowskian spacetime
such as
$so(2,2)$~\cite{beyond,herranz00a}, 
$so(3,2)$~\cite{herranz97}, and more recently   
$so(4,2)$~\cite{Herranz:2002fe,Lukierskidd,Aizawa,Lukierskid}.
From a more general mathematical perspective, twisted  
$so(p,q)$ algebras have been deduced by applying  different
types of Drinfeld's twists
\cite{olmo99,kulish00} but most of them are written in
terms of
 Cartan--Weyl bases.

In spite of all the mathematical (Hopf structures)
background
 so obtained, most of the possible physical applications 
have  mainly  been studied in relation with 
difference-differential symmetries on   uniform
discretizations of the Minkowskian spacetime along a
certain direction, for which the deformation parameter
plays the role of the lattice step
\cite{herranz00a,Herranz:2002fe, Aizawa}. The aim of this
paper is to explore and extract the physical consequences  
provided by  a new   DSR proposal by considering the most
manageable deformation of the Weyl--Poincar\'e algebra,
$U_\tau({\cal WP})$, which arises within a recently
introduced quantum 
  $so(4,2)$ algebra 
\cite{Herranz:2002fe}.    The   Hopf structure and the
deformed momentum representation for $U_\tau({\cal WP})$
are presented in the next section.  Afterwards, by means of
the   usual techniques developed in studying previous DSR
theories,  we deduce   the physical implications. Among
them, finite  boost transformations are obtained in section
3, and a detailed analysis of the range of  the boost
parameter is performed in section 4. We find that the range
of the rapidity depends on the sign of the deformation
parameter and, in general, the behaviour of the rapidity,  
energy and momentum differs from DSR theories based on 
$\kappa$-Poincar\'e.  For a positive deformation parameter,
the situation is rather similar to the undeformed case,
while for a negative value, the range of the rapidity is
restricted between two asymptotic values  for the energy;
furthermore such values determine a  maximum momentum that
depends not only  on the deformation parameter (as in
$\kappa$-Poincar\'e) but also on the deformed mass of the
particle. In     section 5, two proposals for position and
velocity operators are presented: one of them provides a
variable speed of light for massless particles, while the
other one gives rise to a fixed speed of light and conveys
a   new type of generalized uncertainty principle as well.
Some conclusions and remarks close the paper.

%%%%%%%%%%%%%%%%%%%%%%%%%%%%%%%%%%%%%%%%%%%%%%%%%%%%%%%%%
\sect{Quantum Weyl--Poincar\'e algebra}

Let us consider the  mass-like quantum conformal Minkowskian
algebra introduced in \cite{Herranz:2002fe}, 
$U_\tau(so(4,2))$, where $\tau$   is the deformation
parameter.  The ten Poincar\'e generators together with the
dilation  close a Weyl--Poincar\'e (or similitude) Hopf
subalgebra, $U_\tau({\cal WP})\subset U_\tau(so(4,2))$.
Such a deformation is the natural extension to 3+1
dimensions of the results previously presented in
\cite{herranz00a}  for the time-type quantum $so(2,2)$
algebra and, in fact, is based in the Jordanian twist
(introduced in 
\cite{ogiev93}) that underlies the non-standard (or
$h$-deformation)
$U_\tau(sl(2,\R))$ \cite{drinfeld87}. Thus
  $U_\tau(so(4,2))$ verifies the following sequence of 
 Hopf subalgebra   embeddings:
$$
\begin{array}{rll}  U_\tau(sl(2,\R)) \simeq
U_\tau(so(2,1))\subset U_\tau(so(2,2))\!\!&\subset
U_\tau(so(3,2))&\!\!\subset U_\tau(so(4,2)) \\
\cup\qquad &\qquad\quad\cup  &\qquad\quad\cup \\
U_\tau({\cal WP}_{1+1})\!\!&\subset U_\tau({\cal
WP}_{2+1})&\!\!\subset U_\tau({\cal WP}_{3+1})
\end{array} 
$$ This, in turn,   ensures that properties associated to
deformations in  low dimensions are fulfilled, by
construction, in higher dimensions and moreover  any
physical consequence derived from the structure of
$U_\tau({\cal WP})$ is consistent with a full quantum
conformal symmetry that can further be  developed.
Therefore, throughout the paper we will restrict ourselves
to analyse the  DSR theory provided by the  deformed
Weyl--Poincar\'e symmetries after the identification of the
deformation parameter with the Planck length: 
$\tau\sim  L_p$.

 If $\{J_i, P_\mu=(P_0,\> P), K_i,   D\}$ denote the
generators of rotations, time and space translations,
boosts  and dilations, the  non-vanishing deformed
commutation rules and coproduct of   $U_\tau({\cal WP})$ 
are given by~\cite{Herranz:2002fe}:
 \beq
\begin{array}{lll} [J_i,J_j]= \ie\, \varepsilon_{ijk}J_k 
&\qquad [J_i,K_j]=\ie\, \varepsilon_{ijk}K_k  &\qquad
[J_i,P_j]=\ie\,
\varepsilon_{ijk}P_k   \\[2pt] [K_i,K_j]=-\ie\,
\varepsilon_{ijk}J_k  
 &\qquad   [K_i,P_0]=\ie\, {\rm e}^{-\tau P_0} P_i  &\qquad
 [D,P_i]=\ie\,P_i  \\[2pt]
   \displaystyle{ [K_i,P_i]=  \ie\,
   \frac{{\rm e}^{\tau P_0}-1}{\tau} }  &\qquad
\displaystyle{[D,P_0]=\ie\, \frac{1-{\rm e}^{-\tau
P_0}}{\tau}}  &
\end{array}
\label{ba}
\eeq  
 \beq
 \begin{array}{l}
\Delta(P_0)=1\otimes P_0 + P_0\otimes 1   
\qquad \Delta(P_i)=1\otimes P_i + P_i\otimes {\rm e}^{\tau
P_0} 
 \\[4pt]
\Delta(J_i)= 1\otimes J_i + J_i\otimes 1  \qquad\ \
\Delta(D)=1\otimes D + D\otimes {\rm e}^{-\tau P_0}  \\[4pt]
\Delta(K_i)=1\otimes K_i + K_i\otimes 1-\tau D\otimes  {\rm
 e}^{-\tau P_0}P_i  
 \end{array} 
 \label{bc}
\eeq
 where hereafter  we   assume   $\hbar = c =1$,   sum over
repeated indices,  Latin indices $i,j,k=1,2,3$, while Greek
indices
$\mu,\nu=0,1,2,3$. The generators of $U_\tau({\cal WP})$
are Hermitian operators and $\tau$ is a {\em real}
deformation parameter.

Counit $\epsilon$ and antipode $S$ can  directly be 
deduced from (\ref{ba}) and (\ref{bc}); they read
\ben &&\epsilon(X)=0 \qquad X\in\{J_i, P_\mu, K_i,   D\} 
\qquad\quad
\epsilon(1)=1 \cr  &&S(P_0)=-P_0 \qquad S(P_i)=-P_i\,{\rm
e}^{-\tau P_0} \qquad S(J_i)=-J_i  \label{bbc} \\  &&
S(K_i)=-K_i-\tau D P_i \qquad S(D)=-D\,{\rm e}^{\tau P_0}  
\qquad S(1)=1 .
\nonumber
\een

 The   Poincar\'e sector of   $U_\tau({\cal WP})$ (which
does not close a   Hopf subalgebra due to the coproduct of
the boosts (\ref{bc})) provides one useful operator. If
$P_0$ is considered as the energy of a particle, the
deformation of the quadratic Poincar\'e Casimir
\beq M^2 =\left( \f {{\rm e}^{\tau P_0}-1}   {\tau
}\right)^2- {\> P}^2 
\label{bd}
\eeq  can be assumed as the deformed mass-shell condition  
related to the   rest mass $m$   by  
\beq
\ds m=\f{1}{\tau} \ln \left ( 1+\tau M \right) \qquad
\lim_{\tau\to 0} M  =m .
\label{be}
\eeq   Alternatively, the deformed Poincar\'e Casimir
(\ref{bd}) has been used to obtain a time discretization of
the wave or massless Klein--Gordon  equation with quantum
conformal symmetry 
 once   $\tau$ is identified with the time lattice
constant~\cite{herranz00a,Herranz:2002fe}.

 The operator $M$ allows us to introduce the following 
deformed momentum representation  for the Poincar\'e
generators of (\ref{ba}) in terms of $\>p=(p^1,p^2,p^3)$
for a   spinning   massive particle:
\ben  &&P_0=p^0= \frac 1z \ln\left(1+z\sqrt{M^2+\>p^2}
\right) \qquad J_i=\ie\,\varepsilon_{ijk}\,p^k
\f{\partial}{\partial p^j}+S_i \cr  &&\>P=\>p \qquad  K_i
=\ie\sqrt{M^2+\>p^2}\,\f{\partial}{\partial
p^i}+\varepsilon_{ijk}\,\f{p^j S_k}{M+\sqrt{M^2+\>p^2}}
 \label{bbgg}
\een
 provided that the components of the spin $\>S$ fulfil
$[S_i,S_j]=\ie\,\varepsilon_{ijk}S_k$.

%%%%%%%%%%%%%%%%%%%%%%%%%%%%%%%%%%%%%%%%%%%%%%%%%%%%%%%%%
\sect{Deformed finite boost transformations}

The explicit form of the commutation rules    $[K_i,P_\mu]$
in (\ref{ba}) show that the action of the boost generators
on momentum space is deformed, so   we can expect that the
associated finite boost transformations are also deformed
similarly to the $\kappa$-Poincar\'e case
\cite{Bruno:primo,Lukierski:2002df,Bruno:2002wc}.  

  By taking into account that the   Hopf algebra 
$U_\tau({\cal WP})$ (\ref{ba})--(\ref{bbc}) resembles a
bicrossproduct structure, we  introduce the corresponding 
quantum adjoint action~\cite{Karpacz} as
\beq  {\rm ad}_Y X=-\sum_i S( Y_i^{(1)}) X \, Y_i^{(2)}  
\label{bh}
\eeq  provided that the coproduct of $Y$ is written in
Sweedler's notation as  
$\Delta(Y)=\sum_i Y_i^{(1)}\otimes Y_i^{(2)}$ and $S$ is the
antipode (\ref{bbc}). 

By using (\ref{bh}) it can be checked that 
\beq  {\rm ad}_{K_i} {\cal F}(P_\mu) =[K_i, {\cal F}(P_\mu)]
\label{bi}
\eeq  for any momentum-dependent smooth function ${\cal
F}$. Next, we consider a boost transformation along a
generic direction determined by a unitary vector
$\> u$.  If $P^0_{\mu}=(P^0_0,\>{P}^0)$ are the measurements
performed by the first observer with  rapidity  $\xi=0$ and
$P_\mu=(P_0,\>{P})$ the measurements performed by the
second one with arbitrary $\xi$, the finite boost
transformation associated with
$\>u\por\>K$ is obtained from (\ref{bi}) and reads
\beq  P_\mu=\sum_{n=0}^\infty \frac{1}{n!}\,{\rm
ad}^{n)}_{-i\xi\>u\cdot\>K} P^0_\mu   =\exp\left\{-i\xi\, \>
u\por \> K \right\}P^0_\mu
\exp\left\{i\xi \,\> u\por
\> K \right\}.
\label{ca}
\eeq  Hence the infinitesimal boost transformation is given
by
\beq
\f{\d P_\mu}{\d\xi}=-i[\> u\por \> K,P_\mu].
\label{cb}
\eeq    By substituting (\ref{ba}) on (\ref{cb})  we obtain
a system of coupled differential equations:
\beq
\f{\d P_0}{\d\xi}={\rm e}^{-\tau P_0}\, \> u\por \> P
\qquad\quad
\f{\d \> P}{\d\xi}=\> u\left( \frac{ {\rm e}^{\tau
P_0}-1}{\tau}
\right) 
\label{ccb}
\eeq    which give  rise to   a unique non-linear
differential equation for $P_0$: 
\beq
\f{\d^2 P_0}{\d\xi^2}+\tau \left(\f{\d P_0}{\d\xi} 
\right)^2+\f { {\rm e}^{-\tau P_0}-1}{\tau}=0 .
\label{cc}
\eeq  Therefore  we obtain that
\beq P_0(\xi)= \f{1}{\tau}\ln\left( \f{2 +a_+ {\rm
e}^{\xi}+ a_-  {\rm e}^{-\xi}}{2} \right)  \qquad
\> P(\xi)=\frac{\> u}{2\tau}\left(a_+ {\rm e}^{\xi}- a_- 
{\rm e}^{-\xi}\right)+\> a   
\label{cd}
\eeq  where $a_\pm$ and $\> a$  are  integration constants.
By imposing the initial conditions $P_\mu(0)=P_\mu^0$, we
find that
\beq
 a_\pm=\pm\tau \,\>u\por\>P^0+( {\rm e}^{\tau P_0^0}-1)
\qquad
\>a=\>P^0-\>u(\>u\por\>P^0)  
\label{ce}
\eeq  so that the    deformed finite boost transformations
(\ref{ca}) turn out to be
\beq
\begin{array}{l}
 \ds{ P_0(\xi) =\f{1}{\tau}\ln\left(1+({\rm e}^{\tau
P_0^0}-1)
\cosh\xi+
\tau\,\>{u} \por\>{P}^0 \sinh\xi\right)}    \\[8pt]
\ds{ \>{P}(\xi) = \>{{P}^0}+\>{u} \left
(\>{u}\por\>{P}^0(\cosh
\xi-1)+
\f{1}{\tau}({\rm e}^{\tau P_0^0}-1) \sinh \xi \right)} .  
 \end{array} 
 \label{ch}
\eeq
 It can be checked that the deformed mass-shell  condition
(\ref{bd}) remains invariant under   (\ref{ch}).
Furthermore, this result allows us to deduce the
composition of deformed boost transformations along two
directions characterized by  the unitary vectors
$\>u_1,\>u_2$ such that
$\>u_1\por\>u_2=\cos\theta$.  Starting from the initial
observer with
$P^0_{\mu}$ we consider  a first  transformation
$\>u_1\por\>K$ with rapidity $\xi_1$ to a second observer
which measures
$P_{\mu}$, and next a second transformation  $\>u_2\por\>K$
with boost parameter $\xi_2$  to a third observer which
measures 
$P'_{\mu}$: 
$$
\begin{CD} P_\mu^0 @>{{\ {\xi_1,\, \>u_1\cdot \>K}\ }}>>
 P_\mu=P_\mu(\xi_1;P_\mu^0)@>{{\ {\xi_2,\, \>u_2\cdot \>K}\
}}>> 
P'_\mu=P'_\mu(\xi_2;P_\mu)=P'_\mu(\xi_1,\xi_2;P^0_\mu) .
\end{CD}
$$  Due to the expressions (\ref{bi})--(\ref{ca}) this
sequence   corresponds to 
\beq  P'_\mu= \exp\left\{-i\xi_2\, \> u_2\por \> K \right\}
\exp\left\{-i\xi_1\, \> u_1\por \> K \right\}P^0_\mu
\exp\left\{i\xi_1
\,\> u_1\por \> K \right\}\exp\left\{i\xi_2 \,\> u_2\por
\> K \right\} 
\label{cca}
\eeq  which coincides with the classical Lie group
expression.   Nevertheless, we stress that this is a
consequence of the complete Hopf structure underlying the
definition of the quantum adjoint action (\ref{bh}).

The resulting composition is given by
\beq
\begin{array}{l}
\ds{\!\!\!\!\! P'_0  =\f{1}{\tau}\ln\left\{ 1+({\rm e}^{\tau
P_0^0}-1)( \cosh\xi_1  \cosh\xi_2+\sinh\xi_1 
\sinh\xi_2\cos\theta)\right.}\\[8pt]
\ds{\left.\phantom{{}^{{}^0}}  +\tau\,\>{u}_1
\por\>{P}^0 (\sinh\xi_1 \cosh\xi_2+\cosh\xi_1 
\sinh\xi_2\cos\theta)      
-\tau(\>u_1\cos\theta-\>u_2)\por\>P^0\sinh\xi_2 \right\} 
}  
\\[8pt]
\ds{\!\!\!\!\! \>{P}'  = \>{{P}^0}+
\>{u}_2 \left \{ \phantom{{}^{{}^0}}\!\!\!
\>{u}_1\por\>{P}^0(\cosh\xi_1 
\cosh\xi_2\cos\theta+\sinh\xi_1 
\sinh\xi_2)-\>{u}_2\por\>{P}^0\right\}}\\[8pt]
\ds{ \qquad +
\f{1}{\tau}({\rm e}^{\tau P_0^0}-1) \left\{ 
\phantom{{}^{{}^0}}\!\!\! \>u_2 (\sinh\xi_1 \cosh\xi_2
\cos\theta+\cosh\xi_1 
\sinh\xi_2)+\>n\,\sinh\xi_1  
\right\} }\\[8pt]
\ds{\qquad +\>n\,(\>u_1\por\>P^0)(\cosh\xi_1 -1)-\>u_2
(\>n\por\>P^0)\cosh\xi_2      }
 \end{array} 
 \label{ci}
\eeq  where we have introduced the shorthand notation
$\>n=\>u_1-\>u_2\cos\theta$.

As a straightforward consequence, if both deformed boost
transformations are performed along the same direction 
$\>u_1=\>u_2$, then $\theta=0$ and $\>n=0$, so that
$P'_\mu=P'_\mu(\xi_1+\xi_2;P^0_\mu)$, and thus we obtain 
the additivity of the rapidity in the same way as for
$\kappa$-Poincar\'e~\cite{Bruno:2002wc}.

%%%%%%%%%%%%%%%%%%%%%%%%%%%%%%%%%%%%%%%%%%%%%%%%
\sect{Range of  rapidity, energy and momentum}

The stationary points of the energy can be studied by means
of the derivatives of $P_0(\xi)$. For the sake of
simplicity we shall analyse the $(1+1)$-dimensional  case.

%%%%%%%%%%%%%%%%%%%%%%%%%%%%%%%%%%%%%%%%%%%%%%%%
\subsect{Massive particles}

Let us consider firstly massive particles. If we
particularize the transformations  (\ref{ch}) to 
$P_\mu=(P_0,P_1)$, drop  the
 index 0 in $P_\mu^0$, and introduce the  deformed mass
(\ref{bd}), we find that the deformed finite boost
transformations can be rewritten as
\beq
\begin{array}{l}
\ds{ P_0(\xi)=\f{1}{\tau}\ln \left (1+ 
\tau\sqrt{M^2+P_1^2}\cosh\xi + 
\tau P_1\sinh \xi  \right ) }   \\[8pt]
\ds  { P_1(\xi)= {P_1}\cosh \xi+\sqrt{M^2+P_1^2}\sinh\xi  
}.
 \end{array} 
 \label{da}
\eeq  The zero value for the first derivative of $P_0(\xi)$
gives the following expression for the rapidity
\beq
\tanh\xi_0 = - \f{P_1}{\sqrt{M^2+P_1^2}}  \qquad
\xi_0 =\ln \left ( \f{  M}{  \sqrt{M^2+P_1^2}+P_1}
 \right ) 
\label{db}
\eeq  for which  the energy   takes the value of the
physical rest mass of the particle under consideration and
the momentum vanishes:
\beq
 P_0(\xi_{0})=\f{1}{\tau} \ln\left( 1+{\tau M} \right)=m 
\qquad 
 P_1(\xi_{0})=0.
\label{dc}
\eeq  To establish whether this situation corresponds to   a
minimum of the energy,  as it should be,  we  compute the
second derivative of
$P_0(\xi)$. Thus we obtain
\beq
 \frac{{\rm d}^2 P_0(\xi)}{{\rm d}\xi^2}\biggl|_{\xi=\xi_0}
=\f{M}{1+\tau M} .
\label{dd}
\eeq  Therefore    two different situations arise according
to the sign of the deformation parameter:

\noindent
$\bullet$  If $\tau>0$, $\xi_0$  always determines a
minimum of   the energy.

\noindent
$\bullet$ If $\tau<0$,   $\xi_0$ provides a minimum only if
$\ds M < {1}/{|\tau|}$.

%%%%%%%%%%%%%%%%%%%%%%%%%%%%%%%%%%%%%%%%%%%%%%%%%%
%            FIGURE  1                                     
%%%%%%%%%%%%%%%%%%%%%%%%

%%%%%%%%%%%%%%%%%%%%%%%%%%%%%%%%%%%%%%%%%%%%%%%%%%%%
%%%%%%%%%%%%%%%%%%%% figure1 %%%%%%%%%%%%%%%%%%%%%%%
\begin{figure}[t]
\begin{center}
\begin{picture}(290,155)
\footnotesize{
\put(0,65){\vector(1,0){125}}
\put(80,40){\vector(0,1){110}}
\put(51,65){\circle*{3}}
\put(51,55){\makebox(0,0){$\xi_0$}}
\put(120,55){\makebox(0,0){$\xi$}}
\put(62,144){\makebox(0,0){$P_0(\xi)$}}
\put(80,111){\circle*{3}}
\put(92,111){\makebox(0,0){$P_0$}}
\qbezier(15,140)(45,45)(100,140)
\qbezier[15](51,65)(51,78)(51,92)
\qbezier[15](51,92)(65,92)(79,92)
\put(80,92){\circle*{3}}
\put(92,92){\makebox(0,0){$m$}}

\put(180,65){\vector(1,0){110}}
\put(240,25){\vector(0,1){125}}
\put(222,144){\makebox(0,0){$P_1(\xi)$}}
\put(285,55){\makebox(0,0){$\xi$}}
\put(222,65){\circle*{3}}
\put(226,55){\makebox(0,0){$\xi_0$}}
\put(240,109){\circle*{3}}
\put(252,109){\makebox(0,0){$P_1$}}
\qbezier(210,30)(215,52)(240,110)
\qbezier(240,110)(245,120)(254,145) }
\end{picture}
\\[-20pt]

a) Positive deformation parameter $\tau$.
\\[20pt]

\begin{picture}(290,155)
\footnotesize{
\put(0,65){\vector(1,0){125}}
\put(80,40){\vector(0,1){110}}
\put(60,65){\circle*{3}}
\put(60,55){\makebox(0,0){$\xi_0$}}
\put(92,65){\circle*{3}}
\put(92,55){\makebox(0,0){$\xi_+$}}
\qbezier[40](92,65)(92,105)(92,145)
\put(28,65){\circle*{3}}
\put(28,55){\makebox(0,0){$\xi_-$}}
\qbezier[40](28,65)(28,105)(28,145)
\qbezier(35,145)(60,20)(85,145)
\put(120,55){\makebox(0,0){$\xi$}}
\put(62,144){\makebox(0,0){$P_0(\xi)$}}
\put(80,121){\circle*{3}}
\put(67,121){\makebox(0,0){$P_0$}}
\qbezier[10](60,65)(60,73)(60,82)
\qbezier[10](60,82)(69,82)(79,82)
\put(80,82){\circle*{3}}
\put(86,88){\makebox(0,0){$m$}}

\put(180,90){\vector(1,0){110}}
\put(240,25){\vector(0,1){135}}
\put(222,153){\makebox(0,0){$P_1(\xi)$}}
\put(285,80){\makebox(0,0){$\xi$}}
\put(222,90){\circle*{3}}
\put(255,90){\circle*{3}}
\put(255,80){\makebox(0,0){$\xi_+$}}
\qbezier[40](255,90)(255,120)(255,150)
\put(190,90){\circle*{3}}
\put(198,80){\makebox(0,0){$\xi_-$}}
\qbezier[40](190,90)(190,60)(190,30)
\put(226,80){\makebox(0,0){$\xi_0$}}
\put(240,119){\circle*{3}}
\put(248,114){\makebox(0,0){$P_1$}}
\put(255,150){\circle*{3}}
\put(190,30){\circle*{3}}
\qbezier(190,30) (229,105)(240,119)
\qbezier(240,119) (249,132)(255,150)
\qbezier[10](240, 150)(247 ,150 )(255,150 )
\put(240, 150){\circle*{3}}
\qbezier[35](190,30)( 215,30 )( 240,30 )
\put( 240,30){\circle*{3}}
\put(265,30){\makebox(0,0){$ -P_{1,{\rm max}}$}}
\put(280,150){\makebox(0,0){$ +P_{1,{\rm max}}$}} }
\end{picture}
\\[-15pt]

b)   $\tau<0$,  $M<1/|\tau|$ and
$ P_{1,{\rm max}}=\sqrt{1-\tau^2 M^2}/|\tau|$.

\end{center}
\noindent
\\[-30pt]
\caption{Energy $P_0(\xi)$ and momentum $P_1(\xi)$ for a
massive particle with a positive initial momentum $P_1$
according to the sign of $\tau$. }
\label{figure1}
\end{figure}
%%%%%%%%%%%%%%%%%%%%%%%%%%%%%%%%%%%%%%%%%%%%%%
%    END      FIGURE 1            
%%%%%%%%%%%%%%%%%%%%%%%%%%%%%%%%%%%%%%%%%%%%%%%%%%%%%

Now, by taking into account these cases we  analyse the
range of the boost parameter, energy and momentum.   When
$\tau$ is positive the rapidity $\xi$ can take any real
value and the expressions (\ref{da}) show that both
$P_0(\xi),P_1(\xi)$ are always well defined as well as
unbounded (see figure \ref{figure1}a). In particular, in
the limit $\xi\to +\infty$, we find that
$P_0(\xi)\to +\infty$, $P_1(\xi)\to +\infty$, while if 
$\xi\to -\infty$, then $P_0(\xi)\to +\infty$,
$P_1(\xi)\to -\infty$,   in the same way as in the
undeformed case.

On the contrary, when  $\tau$ is negative   the condition 
$\ds M < {1}/{|\tau|}$ must be fulfilled. Then we   
rewrite the energy as
\beq  P_0(\xi)=\f{1}{|\tau |}\ln\left ( \f{1}{1-|\tau|
\sqrt{M^2+P_1^2}
  \cosh\xi -|\tau| P_1 \sinh\xi} \right )  
\label{de}
\eeq  which is always a  well defined expression. However
there exist two values of the boost parameter which give
asymptotic values for the energy, namely
\beq
\xi_- =\ln \left ( \f{1-\sqrt{1-\tau^2 M^2}}{|\tau|
(\sqrt{M^2+P_1^2}+P_1)}
 \right ) \qquad \xi_+ =\ln \left ( \f{1+\sqrt{1-\tau^2
M^2}}{|\tau|
 (\sqrt{M^2+P_1^2}+P_1)} \right )  
\label{df}
\eeq  which are consistent with the constraint  $\ds M <
{1}/{|\tau|}$. Hence  the   range of the boost parameter is
not the whole real axis but it does have a limited range:
$\xi_-<\xi<\xi_+$. This interval is symmetric with respect
to
$\xi_0$ (\ref{db}):
\beq
\xi_+-\xi_0  = \xi_0-\xi_-= \ln \left ( \f{1+\sqrt{1-\tau^2
M^2}}{|\tau| M} \right )  .
\label{dg}
\eeq  The points $\xi_\pm$ show   an unbounded energy but
provide a maximum momentum:
\beq P_0(\xi_\pm) =+\infty \qquad P_1(\xi_\pm) =\pm
\frac{1}{|\tau|}
\sqrt{1-\tau^2 M^2}  .
\label{dh}
\eeq
  Consequently, whenever $\tau$ is negative and  $\ds M <
{1}/{|\tau|}$, we find that the behaviour of $\xi$,
$P_0(\xi)$ and
$P_1(\xi)$ differs from the classical one as depicted in
figure
\ref{figure1}b;  the momentum saturates in the asymptotic
values for the energy  with a maximum value which is
different for each particle since it depends not only on
the deformation parameter but also on the deformed mass.

 %%%%%%%%%%%%%%%%%%%%%%%%%%%%%%%%%%%%%%%%%%%%%%%%
\subsect{Massless particles}

For massless particles with $M=m=0$ we have that 
  ${{\rm e}^{\tau P_0}-1}=    {\tau }|P_1|$,  so that the
equations (\ref{db})--(\ref{dc}) reduce to
\beq
\tanh\xi_0 = -  {P_1}/|P_1|  \qquad
  P_0(\xi_{0})=0  \qquad 
 P_1(\xi_{0})=0.
\label{di}
\eeq  Although the second derivative (\ref{dd}) vanishes,
the behaviour of massless particles is again deeply
determined by the sign of $\tau$.

\begin{figure}[t]
\begin{center}
\begin{picture}(410,155)
\footnotesize{
\put(0,65){\vector(1,0){102}}
\put(80,50){\vector(0,1){100}}
\put(95,55){\makebox(0,0){$\xi$}}
\put(62,144){\makebox(0,0){$P_0(\xi)$}}
\put(80,115){\circle*{3}}
\put(92,115){\makebox(0,0){$P_0$}}
\qbezier(0,68)(50,70)(100,140)

\put(160,65){\vector(1,0){105}}
\put(225,50){\vector(0,1){100}}
\put(207,144){\makebox(0,0){$P_0(\xi)$}}
\put(263,55){\makebox(0,0){$\xi$}}
\qbezier[40](245,65)(245,105)(245,145)
\put(245,65){\circle*{3}}
\put(245,55){\makebox(0,0){$\xi_+$}}
\put(225,98){\circle*{3}}
\qbezier(160,70)(240,80)  (243,145)
\put(235,95){\makebox(0,0){$P_0$}}

\put(300,65){\vector(1,0){105}}
\put(365,50){\vector(0,1){100}}
\put(347,144){\makebox(0,0){$P_1(\xi)$}}
\put(403,55){\makebox(0,0){$\xi$}}
\put(385,65){\circle*{3}}
\put(385,55){\makebox(0,0){$\xi_+$}}
\qbezier[40](385,65)(385,100)(385,135)
\put(365,99){\circle*{3}}
\put(375,95){\makebox(0,0){$P_1$}}
\qbezier[10]  (365,135)(375,135)(385,135)
\qbezier(300,70)(370,80)  (385,135)
\put(385,135){\circle*{3}}
\put(365,135){\circle*{3}}
\put(348,127){\makebox(0,0){$1/|\tau|$}} }
\end{picture}
\\[-40pt]

a)  $\tau>0$; similar for $P_1(\xi)$.
\qquad\qquad\qquad\qquad\qquad  b) 
$\tau<0$.\qquad\qquad\qquad
\qquad\qquad\qquad

\end{center}
\noindent
\caption{Energy $P_0(\xi)$ and momentum $P_1(\xi)$ for a
massless particle with a positive initial momentum $P_1$
according to the sign of
$\tau$. }
\label{figure2}
\end{figure}
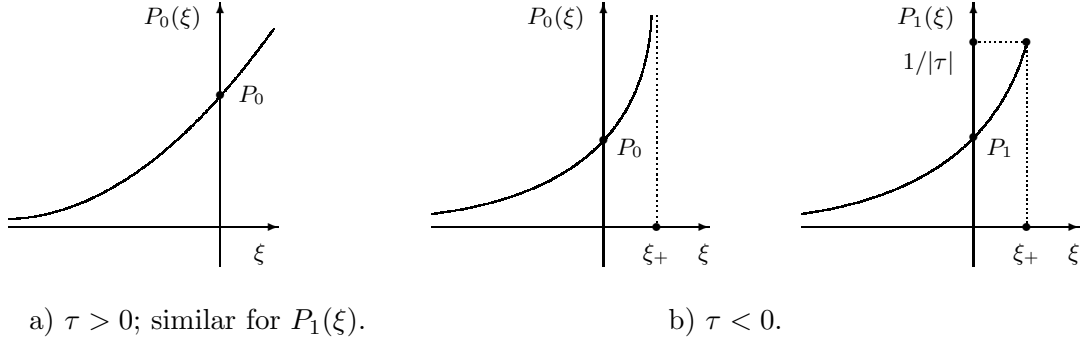

Firstly, let us consider $\tau>0$. If the initial momentum
$P_1>0$, then $\xi_0\to -\infty$ and in the limit
$\xi\to +\infty$, we find that both $P_0(\xi),P_1(\xi)\to
+\infty$. Analogously, if   $P_1<0$, then $\xi_0\to
+\infty$ and
$P_0(\xi)\to +\infty$, $P_1(\xi)\to -\infty$ under the limit
$\xi\to -\infty$. This situation is also rather similar to
the undeformed case (see  figure \ref{figure2}a).

On the other hand, if $\tau<0$, there is no additional
constraint (the above  condition $M=0<1/|\tau|$    is  
trivially fulfilled). According to the sign of the initial
momentum, one of the two previous asymptotes disappears and
coincides with
$\xi_0$. In particular, if $P_1>0$, then
$\xi_0\equiv
\xi_-\to-\infty$, the asymptotic value  $\xi_+$  is left
and the momentum saturates as shown in   figure
\ref{figure2}b:
\beq
\xi\in(-\infty,\xi_+) \qquad  \xi_+ =\ln\left(  
\f{1}{|\tau|
 P_1}  \right)  \qquad P_0(\xi_+)=+\infty \qquad
P_1(\xi_+)=\frac{1}{|\tau|}.
\label{dj}
\eeq  If $P_1<0$, we find that
$\xi_0\equiv
\xi_+\to +\infty$, the rapidity
$\xi_-$ is kept and the momentum also saturates:
 \beq
\xi\in(\xi_-,+\infty) \qquad  \xi_- =-\ln\left(  
\f{1}{|\tau|
 |P_1|} \right)   \qquad P_0(\xi_-)=+\infty \qquad
P_1(\xi_-)=-\frac{1}{|\tau|}.
\label{dk}
\eeq  Therefore, the maximum momentum is only determined by
the deformation parameter, although the corresponding
rapidity
  depends on the initial momentum as well.

%%%%%%%%%%%%%%%%%%%%%%%%%%%%%%%%%%%%%%
%            FIGURE 2          %
%%%%%%%%%%%%%%%%%%%%%%%%%%%%%%%%%%%%%%

%%%%%%%%%%%%%%%%%%%%%%%%%%%%%%%%%%%%%%%%%%%%%%%%%%%%
%%%%%%%%%%%%%%%%%%%% figure2 %%%%%%%%%%%%%%%%%%%%%%%
%%%%%%%%%%%%%%%%%%%%%%%%%%%%%%%%%%%%%%%%%%%%%%
%    END      FIGURE 2    %
%%%%%%%%%%%%%%%%%%%%%%%%%%%%%%%%%%%

%%%%%%%%%%%%%%%%%%%%%%%%%%%%%%%%%%%%%%%%%
\sect{Position and velocity operators}

Position and velocity observables   in DSR theories,
specially for the $\kappa$-Poincar\'e algebra,  have been
introduced from different approaches
\cite{Bacrya,Bacryb,Null,Amelino-Camelia:2002tc,
Kosinski:2002gu,Mignemi:2003ab,Kowalskiotro} and some of
them  lead to different proposals. In our case, we follow
the    same algebraic procedure firstly applied in
\cite{Bacryb} for
  $\kappa$-Poincar\'e and also in  \cite{Null} for the
null-plane quantum Poincar\'e algebra. We consider  some
generic position
$Q_i$ and velocity $V_i$ operators defined as
\beq 
Q_i=\f{1}{2}\left(\f{1}{f(\tau,P_0)}\,K_i+
K_i\,\f{1}{f(\tau,P_0)}
\right) \qquad  V_i=\f{\d Q_i}{\d t}=-\ie [Q_i,P_0] 
\label{ea}
\eeq  where $f(\tau,P_0)$ is an arbitrary smooth function
such that
$\lim_{\tau\to 0}f(\tau,P_0)=P_0$. Next we compute the  
commutation rules between $Q_i$ and the remaining
Weyl--Poincar\'e generators (we omit the arguments in the
function $f$): 
\beq
\begin{array}{l}
\ds{[J_i,Q_j]=\ie\,\varepsilon_{ijk}Q_k \qquad
[Q_i,P_0]=\ie\,
\f{{\rm e}^{-\tau P_0}}{f }\,P_i \qquad  
[Q_i,P_j]=\ie\,\delta_{ij} \, \f{ {\rm e}^{\tau P_0}-1
}{\tau f }   } \\[12pt]
\ds{  [Q_i,Q_j]= \ie\, \f{f' }{f^2 } \,{\rm e}^{-\tau
P_0}\left( 
	Q_iP_j-Q_jP_i-\f{{\rm e}^{\tau P_0}}{f' }
\,\varepsilon_{ijk}J_k
\right)  } \\[14pt]
\ds{[D,Q_i]=-\f{\ie}{2} \left\{ \f{f' }{f^2 } \left(\f{1-
{\rm e}^{-\tau P_0}}{\tau}  
\right)K_i +K_i  \left(\f{1- {\rm e}^{-\tau P_0}}{\tau}  
\right)  \f{f' }{f^2 }
\right\} } 
 \end{array} 
\label{eaa}
\eeq  where $f'$ is the formal derivative of $f$ with
respect to $P_0$. This suggests two natural possibilities
for the function $f$, that are summarized as follows.
\medskip

\noindent $\bullet$  (1) $f =({\rm e}^{\tau
P_0}-1)/\tau=\sqrt{M^2+\>P^2}$. Hence we obtain that
\beq
\begin{array}{l}
\ds{[J_i,Q_j]=\ie\,\varepsilon_{ijk}Q_k \qquad
[Q_i,P_0]=\ie\,
\f{{\rm e}^{-\tau P_0}}{\sqrt{M^2+\>P^2}}\,P_i  \qquad
V_i=\f{{\rm e}^{-\tau P_0}}{\sqrt{M^2+\>P^2}}\,P_i} \\[12pt]
\ds{[D,Q_i]=-\ie\,Q_i \qquad
[Q_i,Q_j]=-\ie\,\varepsilon_{ijk}\,\f{\Sigma_k}{M^2+\>P^2} 
\qquad  [Q_i,P_j]=\ie\,\delta_{ij}} 
 \end{array} 
\label{eb}
\eeq  where we have introduced the kinematical observables 
\cite{Bacrya,Bacryb}:
$\Sigma_k=J_k-\varepsilon_{ijk}Q_iP_j$. By taking into
account the representation (\ref{bbgg}), it can be checked
that  if we consider   a spinless  massive  particle, then
$\>\Sigma=0$ (i.e.\ $\>J=\>Q\times \>P$), so that
$[Q_i,Q_j]=0$,  while  if we consider a spinning massive
particle 
  with $\>P=0$, then $\>\Sigma=\>S$. 

Physical consequences of this choice are directly deduced
from the commutation rules (\ref{eb}): the position $\>Q$
behaves as a classical vector under rotation and dilations,
there is no generalized uncertainty principle and the
velocity of 
  massless particles  is   $|\>V|=V= {\rm e}^{-\tau P_0}$,
so that this  depends on the energy, which can be either 
$V= {\rm e}^{-\tau P_0}<1$ for $\tau>0$, or $V= {\rm
e}^{|\tau| P_0}>1$ for
$\tau<0$. This fact is   well known in DSR theories  based
on
$\kappa$-Poincar\'e
\cite{Bacrya,Amelino-Camelia:1999pm}.

\medskip

\noindent $\bullet$  (2) $f =(1-{\rm e}^{-\tau
P_0})/\tau={\rm e}^{-\tau P_0}\sqrt{M^2+\>P^2}$. In this
case, we find that
\beq
\begin{array}{l}
\ds{[J_i,Q_j]=\ie\,\varepsilon_{ijk}Q_k  \qquad
[Q_i,P_0]=\ie\,
\f{P_i}{\sqrt{M^2+\>P^2}} \qquad
 V_i=\f{ P_i}{\sqrt{M^2+\>P^2}} } \\[12pt]
\ds{[D,Q_i]=-\ie\,\left( {\rm e}^{-\tau P_0} Q_i+Q_i {\rm
e}^{-\tau P_0}
\right) \qquad [Q_i,P_j]=\ie\,\delta_{ij}(1+\tau
\sqrt{M^2+\>P^2}) } \\[8pt]
\ds{ [Q_i,Q_j]=-\ie\,\varepsilon_{ijk}\left(
\f{\Sigma_k}{M^2+\>P^2} +\f{\tau^2 J_k}{\tanh(\tau P_0/2)}
\right)} .
 \end{array} 
\label{ec}
\eeq
 Therefore the position operators are transformed as a
classical vector under rotations, but as a deformed one
under dilations. The velocity for massless particles
reduces to  $V=1$ as in special relativity. In this sense
we remark that, very recently, this result has been
obtained for all known DSR theories in
\cite{Kowalskiotro}, including
$\kappa$-Poincar\'e,  by using a Hamiltonian approach.

Furthermore, if we consider a spinless massive particle,  
the kinematical observables vanish so that position and
momentum operators verify
 \beq
 [Q_i,Q_j]=-\ie\,\varepsilon_{ijk} \,\f{\tau^2
J_k}{\tanh(\tau P_0/2)} \qquad
 [Q_i,P_j]= \ie\,\delta_{ij} {\rm e}^{\tau P_0}  . 
  \label{ed}
\eeq  The latter commutation rule leads to the following
generalized uncertainty principle: 
\beq
\Delta Q_i\Delta P_j\ge \f{1}{2} \delta_{ij}\langle {\rm
e}^{\tau P_0}\rangle=
\f{1}{2} \delta_{ij}\langle 1+\tau \sqrt{M^2+\>P^2}
\rangle 
\label{ee}
\eeq  where $\langle .\rangle$ is the expectation value  and
$\Delta$ here means a root-mean-square deviation.
 We stress that, by following the arguments proposed in 
\cite{Mag},  the expression (\ref{ee}) can be interpreted
as a linear correction in
$\Delta P$ of the usual Heisenberg uncertainty relation,
whilst the
$\kappa$-Poincar\'e construction leads to a quadratic term
in
$\Delta P$. In this respect, see \cite{Garay} for a
comprehensive discussion of deformed uncertainty relations
arising in quantum gravity theories.

%%%%%%%%%%%%%%%%%%%%%%%%%%%%%%%%%%%%%%%%%%%%

\sect{Concluding remarks}

We have presented a   first example of a deformed
relativistic theory    based on a   quantum group symmetry
larger than
 Poincar\'e: the Weyl--Poincar\'e  algebra.  We also expect
that the very same approach may be applied to other quantum
deformations of $\cal WP$ as well as to
  quantum $so(4,2)$ algebras. A first possibility is the so
called `length-like' (or space-type) deformation
$U_\sigma({\cal WP})\subset U_\sigma(so(4,2))$
\cite{Herranz:2002fe}, for which the deformation parameter
has dimensions of length, which  has been shown to be the
symmetry algebra   of a  space discretization  of the
Minkowskian spacetime in one spatial direction. In the
(1+1)-dimensional case
\cite{herranz00a}, there is indeed an algebraic duality
that relates both types of deformations
$U_\tau(so(2,2))\leftrightarrow U_\sigma(so(2,2))$ as well
as $U_\tau({\cal WP}_{1+1})\leftrightarrow U_\sigma({\cal
WP}_{1+1})$, by interchanging the energy $P_0$ and the
momentum $P_1$.  Nevertheless this equivalence is `broken'
in   $3+1$  dimensions in such a manner that   a single
`privileged' discrete space direction arises. Thus,  
isotropy of the  space is removed and this fact may 
preclude further possible physical implications. Another
possibility worth to study  is the twisted conformal algebra
$U_z(so(4,2))$
\cite{Aizawa} which also contains a deformed ${\cal WP}$
subalgebra; this case should provide a DSR theory naturally
adapted to   a null-plane (light-cone) framework,    since
the deformation parameter
$z$ can been interpreted in a natural way as the lattice
step along two null-plane directions.

To end with, we would like to point out that in order to
complete the DSR theory provided by
$U_\tau({\cal WP})$,  the corresponding (dual)   quantum
group should be explicitly computed. This would give rise
to   an  associated non-commutative   Minkowskian spacetime
which, by taking into account  the results of section 4,
should be  different from the well known $\kappa$-Minkowski
spacetime 
\cite{Majid:1994cy,Karpacz,Maslanka,Kowalski-Glikman:2002jr}.
Work on this line is in progress.

%%%%%%%%%%%%%%%%%%%%%%%%%%%%%%%%%%%%%%%%%%%%%
\section*{Acknowledgements}

This work was partially supported  by MCyT and JCyL,  Spain
(Projects BFM2000-1055 and BU04/03), and by INFN-CICyT
(Italy-Spain).

%%%%%%%%%%%%%%%%%%%%%%%%%%%%%%%%%%%%%%%%%%%%%%%%%%%

%%%%%%%%%%%%%%%%%%%%%%%%%%%%%%%%%%%%%%%%%%%%%%%%%%%%%%%
\end{document}